\documentclass[aps,prb,twocolumn,groupedaddress]{revtex4}
\usepackage{amsfonts}
\usepackage{mathrsfs}
\usepackage{graphicx}

\begin{document}

\title{Polaronic quantum diffusion in dynamic localization regime}

\author{Yao Yao\footnote{Electronic address:~\url{yaoyao2016@scut.edu.cn}}}

\affiliation{Department of Physics, South China University of Technology, Guangzhou 510640, China}

\date{\today}

\begin{abstract}
We investigate the quantum dynamics in a disordered electronic lattice with Fibonacci sequence of site energy and off-diagonal electron-phonon coupling within a sub-Ohmic bath by the time-dependent density matrix renormalization group algorithm. It is found that, the slope of the inverse participation ratio versus the coupling strength undergoes a sudden change indicating a transition from the static to the dynamic localization. In the dynamic localization regime, the generated polarons coherently diffuse via hopping-like processes evidenced by the saturated entanglement entropy, providing a novel scenario for the transportation mechanism in strongly disordered systems. The mean-square displacement is revealed to be insensitive to the coupling strength, implying the quantum diffusion behavior survives the energy disorder that prevails in real organic materials.
\end{abstract}


\maketitle

\section{Introduction}

In the physics of localization \cite{review}, a celebrated theory manifests that the wave packet of the electron cannot transport out of the regime scaled by the localization length in disordered systems\cite{Anderson}. This poses a big challenge to reveal the transportation mechanism of both excitons and charge carriers in organic molecules \cite{bredas}. As an important alternative of the silicon-based semiconducting materials, organic molecules suffer from the strong disorder even in its crystalline phase which was supposed to be originated from the thermally-induced intermolecular phonon vibrations \cite{review2}, and the incoherent hopping between molecules was thus regarded to be predominant \cite{review3}. Recent advances of experiment in organic photocells, however, figure out the emergence of the delocalization in the ultrafast charge separation process \cite{Science}, with the statement that an electronic band with delocalized wavefunction serves as the transport channel for the photogenerated charges \cite{Chin3}. Meanwhile, the ubiquitous electron-phonon (e-p) interactions were supposed to play a unique and critical role in suppressing the influence of disorder and producing the transport band, as indicated by both experiments and theories\cite{Science2,Chin4,Ciuchi2,mine1}. The two contrary perspectives of e-p interaction subsequently turn out to be a perfect starting point for clarifying the crossover between the diffusion and delocalization mechanisms.

While talking about the physical origin of localization, Anderson mentioned the lattice deformation which may break the translational invariance and open a large mobility gap in the band \cite{Anderson}. Whereas the investigations of the two issues, disorders and phonons, advanced independently for ages. In organic systems, e.g., the disorders were studied mainly by the kinetic Monte Carlo simulations \cite{KMC} and the phonons were investigated within the framework of small-polaron transformation \cite{Silbey}. Situations changed after Troisi \textit{et al.} proposed the dynamic-disorder prospect in 2006 \cite{Troisi} and the dynamic localization in 2010 \cite{Troisi2}, which established a possible connection between the disorder and the phonons. Afterward, researches bloom in comprehending the charge transport mechanism in organic crystalline materials by taking both issues into consideration \cite{DDM,mine0,DL}. Based on the Ehrenfest dynamics, we have also presented a mechanism to combine the bandlike and hopping transport by introducing the decoherence time into the electron dynamics \cite{mine0}. As a result, in the context of the dynamic localization the vibrational modes are classified into two classes: The low-frequency modes ($<1$ps$^{-1}$) act as the dynamic disorder and the high-frequency modes ($>1$ps$^{-1}$) serve as the source of decoherence \cite{mine0,deco}. In the incoherent regime, the electrons rapidly lose their coherence and stochastically hop among molecules in a kinetic manner like classical particles. Fairly speaking, this incoherent hopping mechanism works well in the transport studies but gets into struggling in elucidating the mechanism of charge generation in photocells stemming from the fact that the binding energy of the charge-transfer state is much larger than the thermal energy, and the latter is traditionally considered to be the driving force of the incoherent hopping.

More recently, Di Sante \textit{et al.} unravelled a novel metal-insulator phase transition considering both the disorder and the e-p interaction in the organic molecules \cite{Ciuchi3,Ciuchi}. They found in the weak-interaction limit of a strongly-disordered system, a mobility gap is opened at the Fermi energy along with the disappearance of the gap of density-of-state. It implies that the formed polaron is mobile in a sense making the strongly disordered system poorly conducting. The suitable parameters for the poor conductor phase are of broad relevance especially with the organic materials, offering an alternative explanation of the charge generation in terms of the quantum diffusion of polarons. In order to see how the polaron moves in the disordered system upon the assistance of phonons, in this paper, we present a full quantum dynamics of the polaron diffusion in a so-called Fibonacci lattice. The dynamic localization picture will be reexamined in a dynamical manner to discuss its applicability in elucidating the mechanism of charge generation in organic solar cells. The paper is organized as follows. In Sec. II the model Hamiltonian and the numerical method are described. The main results are presented in Sec. III and a brief summary is addressed in the final section.

\section{Model and methodology}

We first write down a one-dimensional tight-binding Hamiltonian as ($\hbar=1$)
\begin{eqnarray}
H=\sum_i\epsilon_ic^{\dag}_{i}c_{i}&+&\sum_{\mu}\left[\omega_{\mu}\hat{b}^{\dag}_{\mu}\hat{b}_{\mu}\right.\nonumber\\&+&\left.\gamma_{\mu}(\hat{b}^{\dag}_{\mu}+\hat{b}_{\mu})\cdot\sum_i(c^{\dag}_{i}c_{i+1}+{\rm h.c.})\right],\label{hami}
\end{eqnarray}
where $c^\dagger_{i}(c_{i})$ creates (annihilates) an electron at the $i$-th site with the on-site energy $\epsilon_i$; $\hat{b}^{\dag}_{\mu} (\hat{b}_{\mu})$ is the creation (annihilation) operator of the phonons with the frequency being $\omega_{\mu}$, and $\gamma_{\mu}$ is the off-diagonal e-p coupling strength. Herein, the model consists of two terms, on-site energy of electrons with certain degree of disorder and the phonons off-diagonally coupling to the electron which acts as an equivalent hopping term as that in the normal tight-binding model.

In order to get rid of the numerical uncertainty in generating random on-site energies, a Fibonacci sequence is assigned to $\epsilon_i$ which is generated as follows \cite{Fib1,Fib2}. An energy scale $\epsilon_0$ is firstly defined so as to characterize the degree of disorder, and then the sequences of on-site energy are generated by the recursion formula
\begin{eqnarray}
L_1=\{-\epsilon_0\},~L_2=\{\epsilon_0\},~L_{i+1}=\{L_{i},L_{i-1}\}~(i\geq2).\label{Fib}
\end{eqnarray}
With this procedure of generation, we obtain a series of lattice with quasi-disordered permutation of on-site energy, namely $L_3=\{\epsilon_0,-\epsilon_0\},~~L_4=\{\epsilon_0,-\epsilon_0,\epsilon_0\},~~L_5=\{\epsilon_0,-\epsilon_0,\epsilon_0,\epsilon_0,-\epsilon_0\},\cdot\cdot\cdot$. The number of the lattice site is given by the Fibonacci number defined as $F_{i+1}=F_i+F_{i-1}$, with $F_1=F_2=1$. It has been well-known that in the Fibonacci lattice the electron behaves super-diffusion around the original site with the mean-square displacement (MSD) being proportional to $t^{1.55}$ and $t$ being the time \cite{Fib2}. An electric field will give rise to the localization, with the localization length depending on both the energy disorder and the field \cite{Fib1}. In this work, we do not intend to consider the electric field but merely focus on the e-p coupling, which will also induce the dynamic localization as discussed later. It is also worth noting here that, the utilization of the Fibonacci sequence is mainly aiming at avoiding the statistical averaging and reducing the computational cost. The different generating method of the sequence, such as the Thue-Morse sequence \cite{Fib2}, would give rise to slightly different results in a quantitative way, while the qualitative conclusion drawn in this paper is robust and generic.

Instead of dealing with a single phonon mode \cite{Ciuchi2}, we hereby study a phonon bath coupled to the electrons, with the spectral density being continuous as $J(\omega)=2\pi\alpha\omega^{1-s}_c\omega^{s}{\rm e}^{-\omega/\omega_c}$. We have three parameters in the spectral function, namely the cut-off frequency $\omega_c$, the exponent $s$, and the dimensionless coupling $\alpha$. In practice, $\omega_c$ is taken to be sufficiently large to eliminate any influence on the results. $s$ and $\alpha$ together determine the e-p coupling strength. For simplicity, we fix $s$ to be 0.5, a moderate value for the organic materials \cite{mine1,mine3}. Subsequently, we have totally two free parameters in the system, i.e., $\epsilon_0$ denoting the degree of the disorder and $\alpha$ denoting the e-p coupling strength. It does obviously not matter to fix $\epsilon_0$ to be 0.1 as the energy unit and merely adjust $\alpha$ in the computations.

The orthogonal polynomials adapted time-dependent density matrix renormalization group algorithm \cite{tDMRG,Chin2,Chin1,Guo1,mine2,mine3,mine4} is employed to calculate the dynamics of Hamiltonian (\ref{hami}). Initially the population of the electron is set to localize on the center of the lattice and the phonons are at the ground state. Throughout the work, the total number of site is 34 (the ninth Fibonacci number).

\section{Results and discussions}

\subsection{Dynamic localization}

\begin{figure}
\includegraphics[angle=0,scale=0.43]{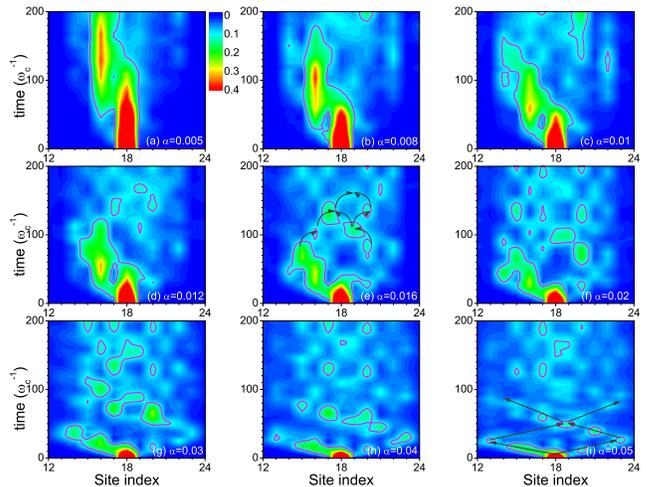}
\caption{Population evolution of electron with nine $\alpha$'s from 0.005 to 0.05. The purple circles denote the populations larger than 0.1. The gray arrows in (e) and (i) denote the hopping directions.}\label{fig1}
\end{figure}

The evolution of electron population is displayed in Fig.~\ref{fig1} for nine values of $\alpha$, with the initial population locating at the site 18 (the original site). For $\alpha=0.005$ and 0.008, the majority of electron population is transferred to site 16 very quickly where the nearest local energy minimum related to the original site is located, implying a polaron is generated therein via the assistance of the e-p coupling, and during a long-term evolution the generated polaron is statically localized and immobile. For $\alpha>0.01$, the situation changes. After a period of time evolution the localized polaron starts to diffuse in a regime, which we follow Troisi \textit{et al.} to call it as the dynamic localization regime \cite{Troisi2,DL}. Taking $\alpha=0.016$ for instance, when the time reaches around 100$\omega_c^{-1}$, some bright spots emerge and are visibly separated indicating the polaron is dynamically localized and the mechanism of the polaronic diffusion is like the hopping mechanism among the sites instead of the continuous wavefunction expansion, i.e., the band transport in a normal electronic lattice. Motivated by the wording ``band-like" and ``band", we hereafter name the new mechanism to be ``hopping-like" mechanism to distinguish from the well-known incoherent hopping mechanism. Following the time advances, the diffused polarons reach the boundary of the dynamic localization regime and turn around and, as indicated by the arrows in Fig.~\ref{fig1}(e), move towards its right side via hopping-like processes. The length between the boundaries of the dynamic localization regime, which is determined merely by the off-diagonal coupling, stands for the dynamic localization length \cite{Troisi2}. We can find that, the stronger the off-diagonal coupling, the longer the dynamic localization length. In addition, more bright spots appear in the dynamic localization regime after long-term evolution, indicating that the polaron has access to more sites. For $\alpha\geq0.03$, the hopping-like mechanism becomes even clearer as figured out by the gray arrows in Fig.~\ref{fig1}(i).

\begin{figure}
\includegraphics[angle=0,scale=1.2]{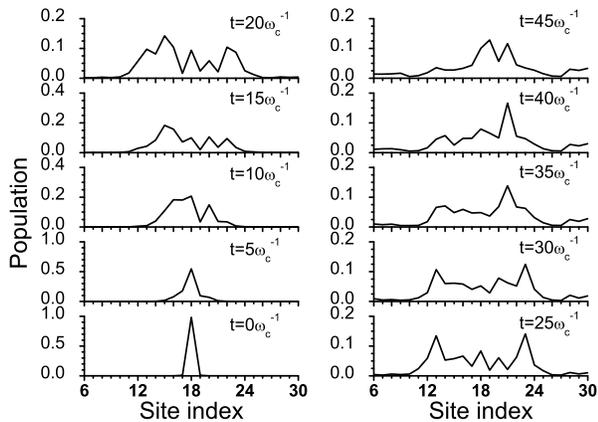}
\caption{Populations of electron at ten time points before $50\omega_c^{-1}$ for $\alpha=0.05$.}\label{fig2}
\end{figure}

To show the polaron diffusion more clearly, the populations of electron at $t<50\omega_c^{-1}$ for $\alpha=0.05$ are displayed in Fig~\ref{fig2}, where the snapshots for every 5$\omega_c^{-1}$ are taken. It is found that the initial single peak of the electron wavepacket residing at the center of the chain splits into two which move individually to each side at $t<25\omega_c^{-1}$. In the snapshot of $t=25\omega_c^{-1}$, one can find two prominent peaks on site 13 and 23, implying that the polaron has moved 5 sites during the period. If the inter-site distance is $a$, the speed of the hopping-like transport is thus 0.2$a\omega_c$. The speed depends on the e-p coupling so that the diffusion is $\alpha$-dependent. At $t>25\omega_c^{-1}$, the two peaks gradually rebound and merge together, implying that site 12 and 24 serves as the boundaries of the dynamic localization regime for $\alpha=0.05$. At $t=40\omega_c^{-1}$, a single peak forming by the oppositely-going peaks emerges at site 21, another energy minimum other than site 16. The process repeats in the subsequent evolution as that shown in Fig.~\ref{fig1}(i).

\subsection{Quantum diffusion}

\begin{figure}
\includegraphics[angle=0,scale=0.32]{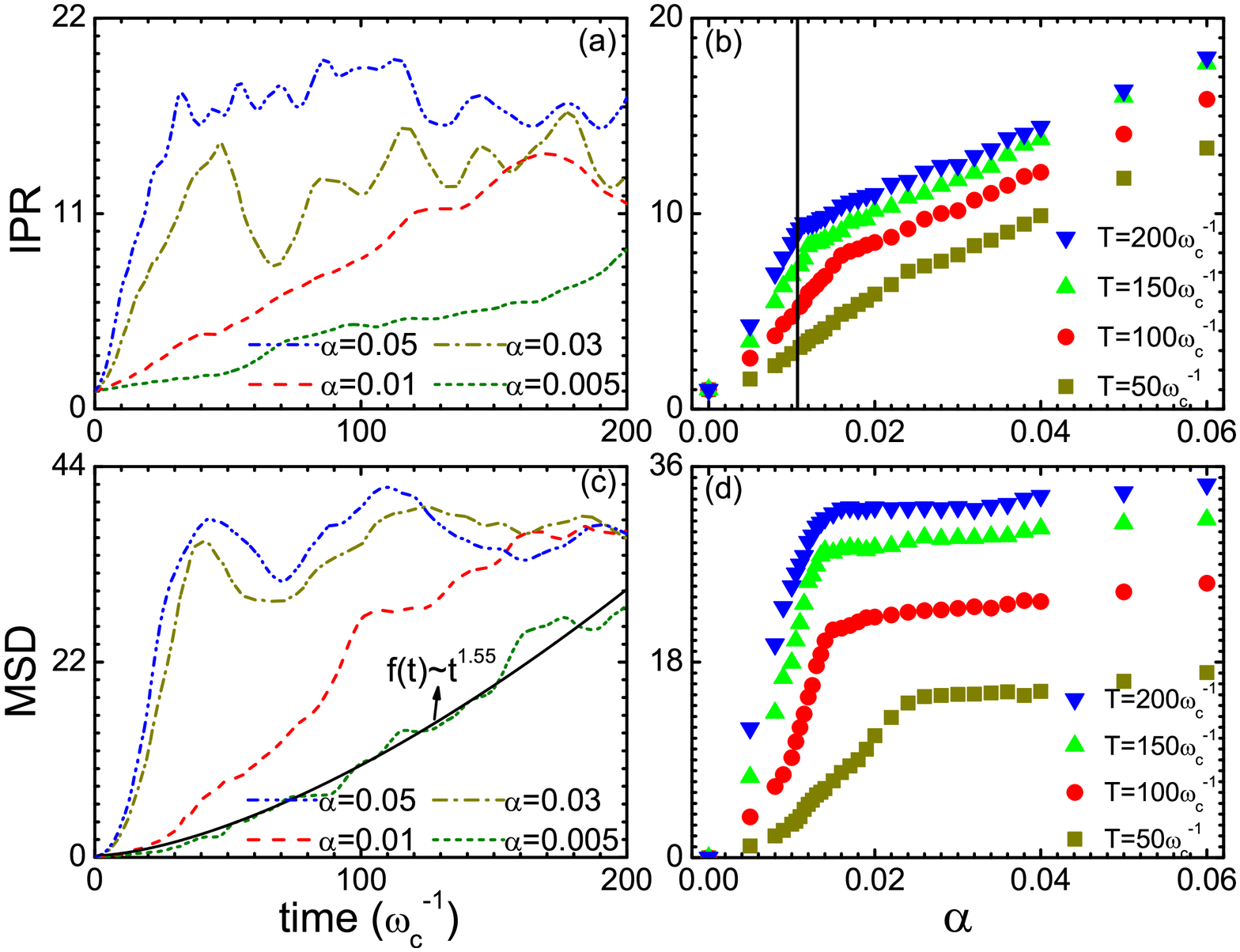}
\caption{(a) Evolution of IPR for four values of $\alpha$. (b) Temporal average value of IPR for four $T$'s versus $\alpha$. The black solid line denotes the transit point of the IPR-$\alpha$ slope. (c) Evolution of MSD for four values of $\alpha$. The black solid curve denotes the fitting function $f(t)\sim t^{1.55}$ characterizing for the super-diffusion. (d) Temporal average value of MSD for four $T$'s versus $\alpha$.}\label{fig3}
\end{figure}

In order to comprehending the transition of static and dynamic localization as well as the hopping-like mechanism, three quantities are calculated on the basis of the electron dynamics. The first one is the inverse participation ratio (IPR) defined by
\begin{eqnarray}
R=\frac{1}{\sum_i\rho_i^2},
\end{eqnarray}
where $\rho_i$ is the electron population at site $i$. IPR is a well-defined measure of the localization length of electron wavefunction. The evolution of IPR is shown in Fig.~\ref{fig3}(a) for four values of $\alpha$. When $\alpha=0.005$, IPR keeps increasing slowly until $t=200\omega_c^{-1}$ implying the polaron extremely slowly gets broadening to the boundary of localization regime. On the other hand, when $\alpha>0.01$ the IPR quickly increases to a saturated value and then oscillates around it. The randomness of the oscillation comes from the energy disorder of the lattice chain. To get insight into the physical meaning of the results we calculate the temporal average via
\begin{eqnarray}
\bar{R}=\frac{1}{T}\int_0^TR\cdot dt.
\end{eqnarray}
The average IPR is displayed in Fig~\ref{fig3}(b) for four $T$'s. When $T=50\omega_c^{-1}$ the average IPR exhibits approximately linear relationship with $\alpha$. On the other hand, when $T>100\omega_c^{-1}$ a turnover emerges at around $\alpha=0.0104$ as indicated by the black solid line in Fig.~\ref{fig3}(b). This sudden change of the IPR-$\alpha$ slope tells us that there is a transition of the polaronic feature from static to dynamic localization. This is equivalent to the phase transition from insulator to poor metal discussed in [20,21], as in the static localization regime the polarons are immobile while in the dynamic localization regime the polarons can move via the quantum diffusion (hopping-like mechanism). The finding is also compatible with the delocalization perspective of charge generation processes in organic solar cells as addressed below.

The second quantity we calculate is the MSD, defined as
\begin{eqnarray}
D=\sum_{i} (i-i_0)^2\rho_i,
\end{eqnarray}
where $i_0$ is the original site. As discussed above, in the absence of electric field the polaron behaves super-diffusion with $D\sim t^{1.55}$.\cite{Fib2} Here, we fit the curve of $\alpha=0.005$ with the function $f(t)\sim t^{1.55}$ and the tendency of the two is found in a good agreement indicating that the very weak coupling case is close to the zero-field one. For larger $\alpha$, the MSD firstly increases and then saturates, sharing the similar tendency with that of IPR. Again, we calculate the temporal average of MSD as shown in Fig.~\ref{fig3}(d). The relationship of MSD and $T$ is similar with that of IPR and $T$. More interestingly, for large $\alpha$ the MSD becomes insensitive to $\alpha$, quite different from the case of IPR. This means for strong coupling the polaron becomes ``larger" than that for weak coupling while the length of the localization regime does not obviously depend on the coupling in the phase of dynamic localization. As one would confuse the ``larger" polaron with the delocalization of electron, this finding is significant to clarify the intrinsic picture of charge generation in organics.

\begin{figure}
\includegraphics[angle=0,scale=0.3]{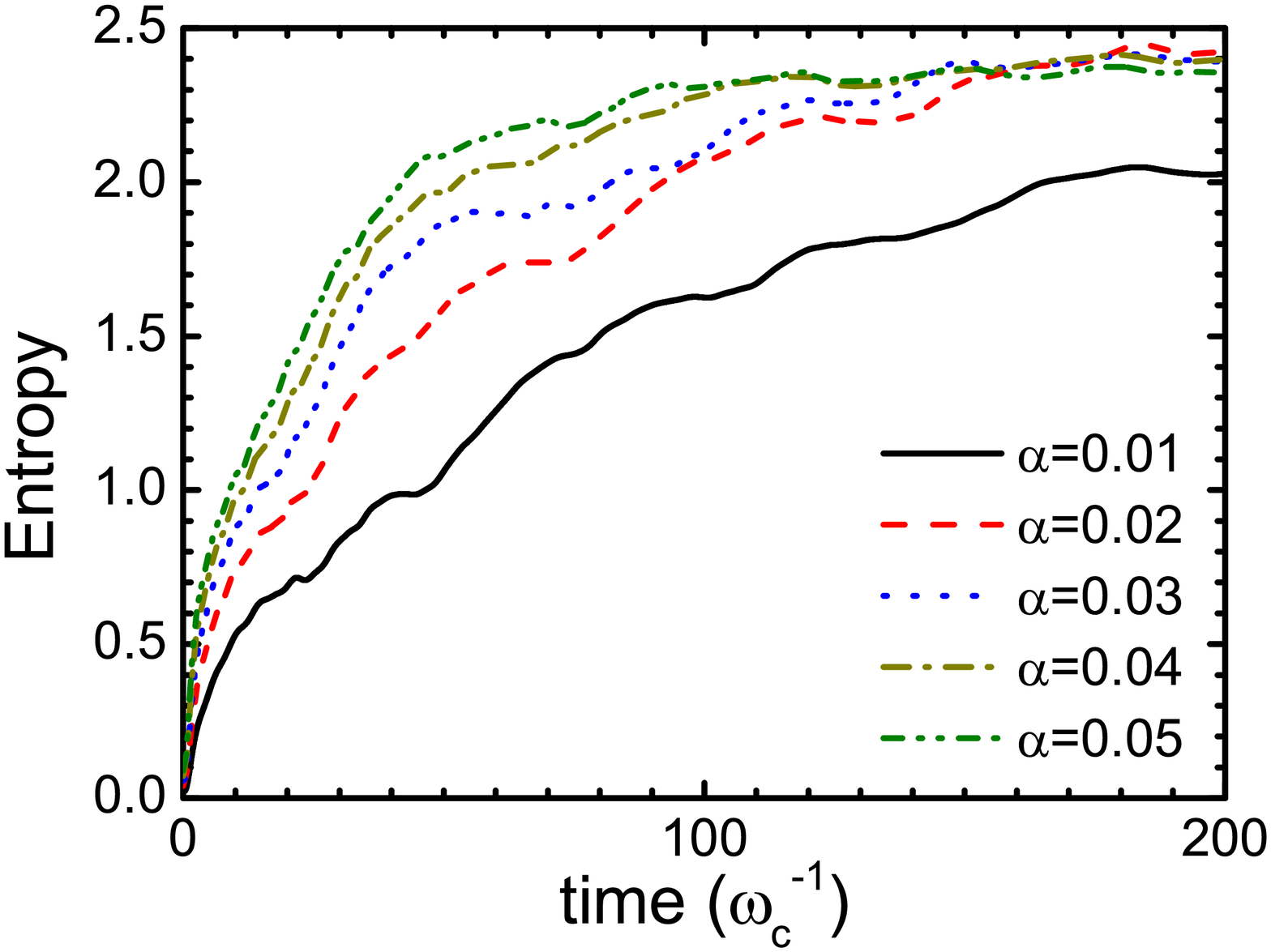}
\caption{Evolution of entropy for five values of $\alpha$.}\label{fig4}
\end{figure}

So far, one would be wondering how the present scenario of polaronic quantum diffusion differs from the classical incoherent hopping mechanism. In other words, is the polaronic diffusion coherent or incoherent? We thus show the evolution of von Neumann entropy of electron in Fig.~\ref{fig4}. The von Neumann entropy is defined as
\begin{eqnarray}
S=-{\rm Tr}\rho\ln\rho,
\end{eqnarray}
where $\rho$ is the reduced density matrix of the electron system. Notable, the entropy is found to quickly increase in the beginning stage of the evolution when the polarons for the first time diffuse to the boundaries of the dynamic localization regime, and then it saturates after $t>100\omega_c^{-1}$ for $\alpha=0.05$. For smaller $\alpha$, it takes longer time for the entropy to saturate. In $t<200\omega_c^{-1}$, the saturated entropies for $\alpha\geq0.02$ converge , and the entropy for $\alpha=0.01$ keeps increasing. The entropy persists in the saturated value during the long-term evolution exhibiting that the quantum coherence is not lost during the polaronic diffusion, otherwise the entropy should decrease significantly since the polaron should incoherently localize in the individual site like the initial situation. As a result, the hopping-like mechanism recognized here is proven to be coherent. Different from the incoherent hopping, it is not assisted by the thermal fluctuation but promoted by the off-diagonal e-p interaction that is accounted for the major driven force of ultrafast long-range charge separation in organic solar cells \cite{mine1}.

\section{Conclusion and outlook}

In summary, we have investigated the polaron dynamics in a Fibonacci lattice with respect to the off-diagonal e-p couplings. Two regimes, static and dynamic localization, are found for weak and strong e-p coupling, respectively. The polaron diffuses quantum-mechanically in the dynamic localization regime, providing a novel scenario of hopping-like mechanism for the charge generation in organic solar cells.

As an outlook, we discuss more about the applicability of the present dynamic localization scenario. In organic materials, the charge photogeneration is accounted to be originated from the wavefunction delocalization \cite{Science,Chin3}, since the electron-spin resonance experiment showed the wavefunction of electron is expanded to an extent of around 10 molecules \cite{length}. Let us now alternatively analyze the current dynamic localization scenario. In common cases, the typical degree of site-energy disorder is 1 to 5$k_BT$, namely 0.026 to 0.12eV at room temperature \cite{Bobbert}. Let the characteristic disorder be 0.1eV, i.e. $\epsilon_0=0.1$ in our model. The turnover value of $\alpha$ from static to dynamic localization as we found is 0.01. Since the typical energy of the intermolecular vibration mode is 7meV, the coupling is then about 25meV, a reasonable coupling strength in organic materials \cite{ep0,ep1,ep2,ep3}. When the coupling is stronger than 25meV, it is inferred from our findings that the charge generation in organic solar cells, which consist of disordered molecular materials with moderate nonlocal e-p couplings, can be elucidated by the mechanism of dynamic localization. The localization length would be about 10 times of the intermolecular distance, which has been measured by the experiment \cite{length}. This dynamic localization scenario is compatible with the conventional delocalization picture \cite{Chin3} in a sense that the wavefunction of electron does not localize in a single molecule. The difference between them is, however, the dynamic localization mechanism survives the relatively strong disorder giving a suitable e-p coupling, while the delocalization does not.

\begin{acknowledgments}
The author gratefully acknowledges support from the National Natural Science Foundation of China (Grant Nos.~91333202 and 11574052).
\end{acknowledgments}


\begin{thebibliography}{99}

\bibitem{review} F. Evers and A. D. Mirlin, Rev. Mod. Phys. \textbf{80}, 1355 (2008).

\bibitem{Anderson} P. W. Anderson, Phys. Rev. \textbf{109}, 1492 (1958).


\bibitem{bredas} V. Coropceanu, J. Cornil, D. A. da Silva Filho, Y. Olivier, R. Silbey, and J. -L. Br\'{e}das, Chem. Rev. \textbf{107}, 926 (2007).


\bibitem{review2} S. Kilina, D. Kilin, and S. Tretiak, Chem. Rev. 115, 5929 (2015).

\bibitem{review3} T. M. Clarke and J. R. Durrant, Chem. Rev. \textbf{110}, 6736 (2010).

\bibitem{Science} A. A. Bakulin, A. Rao, V. G. Pavelyev, P. H. M. van Loosdrecht, M. S. Pshenichnikov, D. Niedzialek, J. Cornil, D. Beljonne, and R. H. Friend, Science \textbf{335}, 1340 (2012).

\bibitem{Chin3} S. L. Smith and A. W. Chin, Phys. Chem. Chem. Phys. \textbf{16}, 20305 (2014).

\bibitem{Science2} S. M. Falke, C. A. Rozzi, D. Brida, M. Maiuri, M. Amato, E. Sommer, A. De Sio, A. Rubio, G. Cerullo, E. Molinari, and C. Lienau, Science \textbf{344}, 1001 (2014).

\bibitem{Chin4} S. L. Smith and A. W. Chin, Phys. Rev. B \textbf{91}, 201302(R) (2015).

\bibitem{Ciuchi2} S. Bera, N. Gheeraert, S. Fratini, S. Ciuchi, and S. Florens, Phys. Rev. B \textbf{91}, 041107(R) (2015).

\bibitem{mine1} Y. Yao, X. Xie, and H. Ma, J. Phys. Chem. Lett. \textbf{7}, 4830 (2016).

\bibitem{KMC} H. B\"{a}ssler, phys. stat. sol. (b) \textbf{175}, 15 (1993).

\bibitem{Silbey} D. R. Yarkony and R. Silbey, J. Chem. Phys. \textbf{67}, 5818 (1977).

\bibitem{Troisi} A. Troisi and G. Orlandi, Phys. Rev. Lett. \textbf{96}, 086601 (2006).

\bibitem{Troisi2} A. Troisi, Phys. Rev. B \textbf{82}, 245202 (2010).

\bibitem{DDM}  A. Troisi and D. L. Cheung, J. Chem. Phys. 131, 014703 (2009); S. Fratini and S. Ciuchi, Phys. Rev. Lett. 103, 266601 (2009); S. Ciuchi, S. Fratini, and D. Mayou, Phys. Rev. B 83, 081202(R) (2011); L. Wang, Q. Li, Z. Shuai, L. Chen, and Q. Shi, Phys. Chem. Chem. Phys. 12, 3309 (2010).

\bibitem{mine0} Y. Yao, W. Si, X. Hou, and C. Q. Wu, J. Chem. Phys. \textbf{136}, 234106 (2012).

\bibitem{DL} T. Wang and W. -L. Chan, J. Phys. Chem. Lett. \textbf{5}, 1812 (2014).

\bibitem{deco}  J.-D. Picon, M. N. Bussac, and L. Zuppiroli, Phys. Rev. B \textbf{75}, 235106 (2007).

\bibitem{Ciuchi3} D. Di Sante and S. Ciuchi, Phys. Rev. B \textbf{90}, 075111 (2014).

\bibitem{Ciuchi} D. Di Sante, S. Fratini, Vladimir Dobrosavljevi\'{c}, and S. Ciuchi, arXiv:1604.07816.


\bibitem{Fib1} H. N. Nazareno, P. E. de Brito, and C. A. A. da Silva, Phys. Rev. B 51, 864 (2015).

\bibitem{Fib2}  P. E. de Brito, C. A. A. da Silva, and H. N. Nazareno, Phys. Rev. B 51, 6096 (2015).





\bibitem{tDMRG} S. R. White, Phys. Rev. Lett. \textbf{93}, 076401 (2004).

\bibitem{Chin2} A. W. Chin, \'{A}. Rivas, S. F. Huelga, and M. B. Plenio, J. Math. Phys. \textbf{51}, 092109 (2010).

\bibitem{Chin1} J. Prior, A. W. Chin, S. F. Huelga, and M. B. Plenio, Phys. Rev. Lett. \textbf{105}, 050404 (2010).

\bibitem{Guo1} C. Guo, A. Weichselbaum, S. Kehrein, T. Xiang, and J . von Delft, Phys. Rev. B \textbf{79}, 115137 (2009).

\bibitem{mine2} Y. Yao, L. Duan, Z. L\"{u}, C. Q. Wu, and Y. Zhao, Phys. Rev. E \textbf{88}, 023303 (2013).

\bibitem{mine3} Y. Yao, Phys. Rev. B \textbf{91}, 045421 (2015); \textit{ibid} \textbf{93}, 115426 (2016).

\bibitem{mine4} Y. Zhao, Y. Yao, V. Chernyak, and Y. Zhao, J. Chem. Phys. \textbf{140}, 161105 (2014); Y. Yao, N. Zhou, J. Prior, and Y. Zhao, Sci. Rep. \textbf{5}, 14555 (2015).


\bibitem{Bobbert} R. Coehoorn, W. F. Pasveer, P. A. Bobbert, and M. A. J. Michels, Phys. Rev. B \textbf{72}, 155206 (2005).

\bibitem{ep0} R. C. Hatch, D. L. Huber, and H. H\"{o}chst, Phys. Rev. Lett. \textbf{104}, 047601 (2010).

\bibitem{ep1} A. Girlando, L. Grisanti, and M. Masino, Phys. Rev. B \textbf{82}, 035208 (2010).

\bibitem{ep2} Y. Li, Y. Yi, V. Coropceanu, and J.-L. Br\'{e}das, Phys. Rev. B \textbf{85}, 245201 (2012).

\bibitem{ep3} Y. Li, V. Coropceanu, and J.-L. Br\'{e}das, J. Chem. Phys. \textbf{138}, 204713 (2013).

\bibitem{length}  K. Marumoto, S. Kuroda, T. Takenobu, and Y. Iwasa, Phys. Rev. Lett. \textbf{97}, 256603 (2006).


\end{thebibliography}
\end{document}